\documentclass{article}
\usepackage{graphicx}

\begin{document}
\topskip 2cm
\begin{titlepage}
\begin{center}
{\large
\bf P-P Total Cross Sections at VHE from Accelerator Data} \\
\vspace{2.0cm} {\large J. P\'{e}rez-Peraza, A. S\'{a}nchez-Hertz, M. Alvarez-Madrigal} \\
\vspace{.5cm}
{\em Instituto de Geof\'{\i}sica, UNAM,
Coyoac\'an, M\'{e}xico D.F., MEXICO} \\
\vspace{1cm}
{\large J. Velasco, A. Faus-Golfe} \\
\vspace{.5cm}
{\em IFIC -- Instituto de F\'{\i}sica Corpuscular\\
Centro Mixto CSIC--Universitat de Val\`encia, Spain}\\
\vspace{1cm}
{\large A. Gallegos-Cruz} \\
\vspace{.5cm} {\em Ciencias B\'asicas, UPIICSA, I.P.N.,
Iztacalco, Mexico D.F, MEXICO}\\
\vspace{2.5cm}
\begin{abstract}
Up today estimations of proton-proton total cross sections,
$\sigma_{tot}^{pp}$, at very high energies are obtained  from
cosmic rays  ($ > 10^{17}$ eV) by means of some approximations and
the knowledge of the measured proton-air cross section at these
energies. Besides, $\sigma_{tot}^{pp}$ are measured   with present
day high energy colliders up to nearly $2$ TeV in the center of
mass ($\sim\ 10^{15}$ eV in the laboratory). Several theoretical,
empirical or semi-empirical  parameterizations, very successful
for interpolation at accelerator  energies, can then be used to
extrapolate the measured value to cosmic ray energies and get a
reasonable estimation of cross sections at higher energies ($\sim\
10^{17}$ eV).  When both cross section estimations - from
accelerators   and cosmic rays - are compared, a disagreement  is
observed, amounting to more than $10 \% $, showing a discrepancy
beyond statistical errors. Here we use a phenomenological model
based on the Multiple-Diffraction  approach to successfully
describe data at accelerator energies. Then we estimate with this
model $\sigma_{tot}^{pp}$ at cosmic ray energies. The employed
fitting method associates the model free-parameters to the data of
only two physical observable: the differential cross-section and
the parameter $\rho$. The model predictions for estimations of
$\sigma_{tot}^{pp}$ are then compared with data of total
cross-sections. On the basis of regression analysis we determine
confident error bands, analyzing the sensitivity of our
predictions to the employed data for extrapolation. : using data
at 546 and 1.8 TeV, our extrapolations for $\sigma_{tot}^{pp}$ are
only compatible with  the Akeno cosmic ray data, predicting a
slower rise with energy than other cosmic ray results and other
extrapolation methods. We discuss our results within the context
of constraints in the light of future accelerator and cosmic ray
experimental results.
\end{abstract}
\vspace{5cm}
\end{center}
\end{titlepage}
\newpage
\setcounter{page}{1}
\renewcommand{\thepage}{\arabic{page}}
\section*{INTRODUCTION}
Recently a number of difficulties in uniting accelerator
and cosmic ray values of  proton-proton, $\sigma_{tot}^{pp}$,
and  antiproton-proton, $\sigma_{tot}^{\bar{p}p}$,
hadronic total  cross-sections,
 within the frame
of the highest up-to-date data have been summarized \cite{Engel1}.
Such united picture appears to be of the highest importance for
the interpretation of results of new cosmic ray experiments as the
HiRes \cite{hires} and in designing proposals that are currently
in progress as the Auger Observatory \cite{Auger}, as well as in
designing detectors for future accelerators, as the CERN pp Large
Hadron Collider (LHC) \cite{Lipari}. Although most of accelerator
measurements of $\sigma_{tot}^{pp}$ and $\sigma_{tot}^{\bar{p}p}$
at center of mass energies $\sqrt{s} \leq $ 1.8 TeV are quite
consistent among them, this is unfortunately not the case for
cosmic ray experiments at $\sqrt{s} > 6 $ TeV where some
disagreements exist among different experiments. This is also the
case among different predictions from the extrapolation of
accelerator data up to cosmic ray energies: whereas some works
predict smaller values of $\sigma_{tot}^{pp}$ than those of cosmic
ray experiments (e.g. \cite{DL1,Augier1}) other predictions  agree
at some specific energies with cosmic ray results (e.g.
\cite{MM1}). Dispersion of cosmic ray results are mainly
associated to the strong model-dependence of the relation between
the basic hadron-hadron cross-section and the hadronic
cross-section in air. The later determines the attenuation  length
of hadrons in the atmosphere, which is usually measured in
different ways, and depends strongly on the rate ($k$) of energy
dissipation of the primary proton into the electromagnetic shower
observed in the experiment: such a cascade is simulated by
different Monte Carlo technics implying additional discrepancies
between different experiments. Furthermore, $\sigma_{tot}^{pp}$ in
cosmic ray experiments is determined from $\sigma_{p-air}^{inel}$
using a nucleon-nucleon scattering amplitude which is frequently
in disagreement with most of accelerator data \cite{Engel1}.

On the other hand, we dispose of many parameterizations (purely
theoretically, empirical or semi-empirical based) that fit pretty
well the accelerator  data. Most of them agree that at the energy
of the future LHC (14 TeV in the center of mass) or higher the
rise in energy of $\sigma_{tot}^{pp}$ will continue, though the
predicted values differ from model to model. We claim that both
the cosmic ray and parametrization approaches must complement each
other in order to draw the best description of the proton-proton
hadronic cross-section behavior at ultra high energies. However,
the present status is that due to the fact that interpolation of
accelerator data is nicely obtained with most of parametrization
models, it is expected that their extrapolation to higher energies
be highly confident: as a matter of fact, parameterizations are
usually based in a short number of fundamental  parameters, in
contrast with the difficulties found in deriving
$\sigma_{tot}^{pp}$ from cosmic ray results \cite{Engel1}. With
the aim of contributing to the understanding of this problem, in
this paper we first briefly analyze in the first two sections the
way estimations are done  for  $\sigma_{tot}^{pp}$ from
accelerators as well as from cosmic rays. We find serious
discrepancies among both estimations. In section III we describe
the Multiple Diffraction model and the employed method to evaluate
the model parameters on basis of data of only two physical
observable: the differential cross-section and the parameter
$\rho$. For the goal of the present approach we neglect crossing
symmetry. In the fourth section we present a suitable
parametrization to high energies of the free energy-dependent
parameters of the model and we discuss its physical significance.
In the fifth section we describe the procedure for the
determination of error bands on basis of Appendix A. In  section
VI, on the basis of the Multiple Diffraction model applied to
accelerator data, we predict $\sigma_{tot}^{pp}$ values with
highly confident errors. We discuss in section VII our results in
terms of the hypothesis $\sigma_{tot}^{\bar{p}p} =
\sigma_{tot}^{pp}$ , and finally, we conclude with a discussion
about the implications of extrapolations within the frame of
present cosmic ray estimations.

\section{HADRONIC  $\sigma_{tot}^{pp}$  FROM ACCELERATORS}
Since the first results of the Intersecting Storage Rings(ISR) at
CERN arrived in the 70s, it is a well established fact that
$\sigma_{tot}^{pp}$ rises with energy (\cite{Amaldi1,ISR1}). The
CERN $S\bar{p}pS$ Collider found this rising valid for
$\sigma_{tot}^{\bar{p}p}$ as  well \cite{UA41}. Later, the
Tevatron at  Fermilab confirmed that for $\sigma_{tot}^{\bar{p}p}$
the rising still continues at 1.8 TeV, even if there is a
disagreement among the  different experiments as for the exact
value \cite{E710,CDF,Avil}. A full discussion on these problems
may be found in  \cite{GM2,Blois99}. It remains now to estimate
the amount of rising of the total cross section at those energies.
Let us resume the standard technique used by accelerator
experimentalists \cite{Augier1}.

Using  a semi-empirical parametrization based on Regge theory and
asymptotic theorems experimentalists have successively described
their data from the ISR to the $S\bar{p}pS$ energies. It takes
into account  all the available data  for $\sigma_{tot}^{pp}$,
$\rho^{pp}$, $\sigma_{tot}^{\bar{p}p}$ and $\rho^{\bar{p}p}$,
where  $\rho^{pp,\bar{p}p}$, is the  ratio of the  real to the
imaginary part of the ($pp, \bar{p}p$) forward elastic amplitude
at $t=0$. The fits are performed using the once-subtracted
dispersion relations:
\begin{equation}
\rho_{\pm}(E)\sigma_{\pm}(E)  \!\! = \!\!  {{C_{s}}\over {p}}+
{{E}\over {\pi p}} \int_{m}^{\infty} dE' p' \{
{{\sigma_{\pm}(E')}\over {E'(E'-E)}}-
 {{\sigma_{\mp}(E')}\over {E'(E'+E)}} \}
\end{equation}
where $C_{s}$ is the substraction constant.
The expression for $\sigma_{tot}^{pp,\bar{p}p}$ is:
\begin{equation}
\sigma_{-,+}^{tot} = A_{1}E^{-N_{1}} \pm A_{2}E^{-N_{2}} + C_{0} +
                    C_{2}[ln(s/s_{0})]^{2}
\end{equation}
where -  (+)  stands for $pp$ ($\bar{p}p$)  scattering. Cross
sections are measured in mb and energy in GeV, $E$ being the
energy measured in the lab frame. The scale factor $s_{0}$ has
been arbitrarily chosen equal to 1 GeV$^{2}$.  The most
interesting piece is the one controlling the high-energy behavior,
given by a $ln^{2}(s)$ term, in order to be compatible,
asymptotically, with the Froissart-Martin bound \cite{FM1}.

The parametrization assumes
$\sigma_{tot}^{pp}$ and $\sigma_{tot}^{\bar{p}p}$ to be the same
asymptotically. This is justified  from the very precise
measurement of the $\rho_{\bar{p}p}$ parameter  at 546 GeV
at the $S\bar{p}pS$ collider,
$\rho_{\bar{p}p} = 0.135 \pm 0.015$ \cite{Augier2},
which implies that at present
there is no sizeable contribution  of the odd under  crossing part of the
forward  amplitude, the  so-called ``Odderon hypothesis''.
This hypothesis predicts a value of $\rho_{\bar{p}p} > 0.17 - 0.20$
\cite{LN}, \cite{GLN}.

The eight free parameters are determined by a fit which minimizes
the $X^{2}$ function
\begin{equation}
X^{2} = X^{2}_{\sigma_{\bar{p}p}}+X^{2}_{\rho_{\bar{p}p}}
            + X^{2}_{\sigma_{pp}}+X^{2}_{\rho_{pp}}
\end{equation}
The fit has proved its validity predicting from the ISR $pp$ and
$\bar{p}p$data
(ranging from 23  to 63 GeV in the center of mass), the
$\sigma_{tot}^{\bar{p}p}$ value \cite{Amaldi2}
later found at the $S\bar{p}pS$ Collider,
one order of magnitude higher in energy (546 GeV)
\cite{UA41}. With the same well-known method and using
the most recent results it is possible to get estimations
for $\sigma_{tot}^{pp}$ at the  LHC and higher  energies. These estimations,
together with our present experimental
knowledge for both $\sigma_{tot}^{pp}$ and $\sigma_{tot}^{\bar{p}p}$
are plotted in figure 1. We have also plotted the cosmic
ray experimental data from AKENO (now AGASSA) \cite{Akeno1} and
the Fly's Eye experiment \cite{FE1,FE2}. The curve is
the result of the fit described in \cite{Augier1}. The increase in
$\sigma_{tot}^{pp}$ as the energy increases is clearly seen.

\begin{figure}
\centering
\includegraphics[width=8cm]{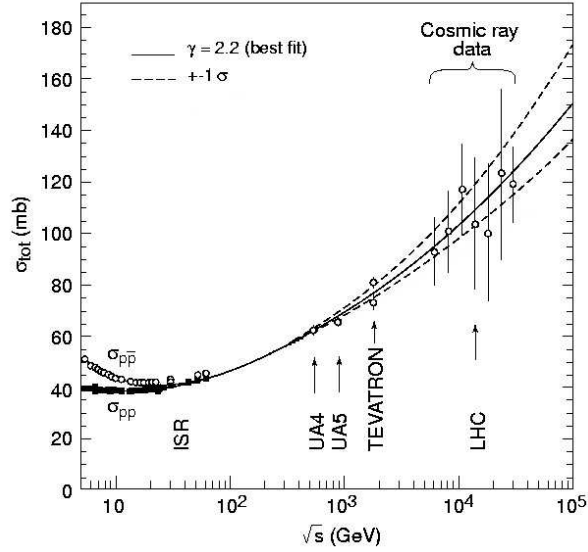}
\caption{$\sigma_{exp}^{tot}$ from accelerator and from cosmic
rays: solid line indicates the best $X^{2}$ fit obtained from a
semi-empirical parametrization \cite{Augier1}. The two dashed
lines delimitate the region of uncertainty. [ Cosmic ray
estimations after \cite{Akeno1}. }
\end{figure}

\begin{table}
\centering
\caption{$\sigma_{tot}^{\bar{p}p}$ data from high
energy accelerators: fits values are from \cite{Augier1}.}
\vspace{1cm}
\begin{tabular}{rlr}
\hline \hline
$\sqrt{s}$  (TeV)& & $ \sigma_{tot}^{pp}$  (mb) \\
\hline
0.55 & Fit  &  $61.8  \pm 0.7 $\\
     & UA4  &  $62.2  \pm 1.5 $\\
     & CDF  &  $61.5  \pm 1.0 $\\
1.8  & Fit  &  $76.5  \pm 2.3 $\\
     & E710 &  $72.8  \pm 3.1 $\\
     & CDF  &  $80.3  \pm 2.3 $\\
     & E811 &  $71.7  \pm  2.0 $\\
16.0   & Fit  & $111.0  \pm 8.0 $\\
40.0   & Fit  & $130.0 \pm 13.0 $\\
\hline \hline
\end{tabular}
\end{table}

Numerical predictions  from this analysis are given  in Table 1.
It should be remarked that at the LHC energies and beyond  the predicted
$\sigma_{tot}^{pp}$ and $\sigma_{tot}^{\bar{p}p}$ values  from
the fitting results display relatively high error values
($\Delta\sigma^{tot}_{pred} \geq  8 mb$).
Also, recent extrapolations for calculations of error bands,
based on analysis  taking  into account cosmic  ray  and  accelerator data,
found error values relatively high,
as for instance 5 - 7 mb at 14 TeV  \cite{Luna}.
{\em We conclude  that it is necessary to look for ways to reduce the  errors
and make the extrapolations more precise}. \\

\section{HADRONIC $\sigma_{\lowercase{tot}}^{\lowercase{pp}}$ FROM COSMIC RAYS}
Cosmic ray experiments give us  $\sigma_{tot}^{pp}$  in an indirect way:
we  have to derive it from cosmic ray extensive air
shower (EAS) data. But, as summarized in \cite{Engel1} and widely discussed
in the literature, the determination of
$\sigma_{tot}^{pp}$ is a rather complicated process with at least two well
differentiated steps. In the first  place,  the primary
interaction involved in EAS is  proton-air; what it is determined through EAS
is the $p$-inelastic cross section,
$\sigma_{inel}^{p-air}$, through some measure of the attenuation of the rate
of showers, $\Lambda_{m}$,  deep in the atmosphere:
\begin{equation}
\Lambda_{m} = k \lambda_{p-air}  = k \ \frac {14.5 m_{p}}
              {\sigma_{inel}^{p-air}}
\end{equation}
The  $k$ factor parameterizes the rate at which the energy of the primary
proton is dissipated into electromagnetic energy.
A simulation with a full representation of the hadronic interactions
in the cascade is needed to calculate it. This is done by
means of Monte Carlo simulations \cite{Pryke,Hillas,Fletcher}. Secondly,
the connection between $\sigma_{inel}^{p-air}$
and $\sigma_{tot}^{pp}$ is model dependent. A theory for nuclei interactions
must be used. Usually is Glauber's theory
\cite{G1,GM1}. The whole procedure makes hard to get a general agreed value
for $\sigma_{tot}^{pp}$. Depending on the
particular assumptions made the values may oscillate by  large amounts,
from as low as $122 \pm 11$  at $\sqrt{s} = 30$ TeV quoted
by the Fly's Eye  group \cite{FE1,FE2} or
$133 \pm 10$ mb by the Akeno Collaboration \cite{Akeno1}, both
at $\sqrt{s} = 30$, to
 $162 \pm 38$ mb  nearly $\sqrt{s} = 30$ ( around $160 - 170$ mb  $\sqrt{s} = 40 $ TeV), \cite{Niko}
 and even as high as  $175_{-27}^{+40}$also at
$\sqrt{s} = 40 $ TeV  \cite{GSY}  .  In the last  cases, and even taking  into
account the large  quoted  errors, the values for
$\sigma_{tot}^{pp}$ are  hardly compatible with the values
obtained  from the extrapolations of current accelerator data.

From this analysis the conclusion is that cosmic ray estimations
of $\sigma_{tot}^{pp}$ are not of much help to constrain
extrapolations from accelerator energies.
Conversely we could ask if a trustable extrapolation based  on  accelerator
data could not be used
to constrain  cosmic ray estimations.

\section{A MULTIPLE-DIFFRACTION APPROACH FOR EVALUATION
OF $\sigma_{tot}^{pp}$ }
Let us tackle the mismatching between accelerator and cosmic ray estimations
using the multiple-diffraction model applied  to  hadron-hadron scattering
\cite{GV1}. The elastic hadronic scattering amplitude for the collision
of two hadrons  A and B, neglecting spin,  is described  as
\begin{equation}
F(q,s) = i \int_{0}^{\infty } b \ db \left[ 1 - e^{i \xi (b,s)}
 \right]  J_{o}(qb)
\end{equation}
where $\xi (b,s)$ is the eikonal, $b$ the impact parameter,
$J_{o}$ the zero-order Bessel function and $q^{2} = - t$ the
four-momentum transfer squared. In the first Multiple Diffraction
Theory the eikonal in the transferred momentum space is
proportional to the product of the hadronic form factors $G_{A}$
and $G_{B}$ (geometry) and the averaged elementary scattering
amplitude among the constituent partons $f$ (dynamics), and can be
expressed at first order as $\xi (b,s) = C_{A,B}\left< G_{A}G_{B}
f \right>$, where the proportional factor $C_{A,B}$ is the free
parameter known as the {\it absorption factor} and the brackets
denote the symmetrical two-dimensional Fourier transform. A
connection between theory and experiment may be obtained by means
of hadronic factors and elementary parton-parton amplitudes, which
are not physical observable. However with the help of the optical
theorem $\sigma_{tot}^{pp}$ may be evaluated in terms of  the
elastic amplitude $F(q,s)$:

\begin{equation}
\sigma_{tot}^{pp} = 4 \ \pi \ ImF(q=0, s)
\end{equation}
$\sigma_{tot}^{pp}$ is a physical observable, the other two
physical observable are the differential cross section and $\rho$
expressed respectively as
\begin{equation}
{d\sigma\over{dq^{2}}}  =  {\pi\mid{F(q,s)}\mid^2}
\end{equation}
and
\begin{equation}
{\rho} =  {\ ReF(q=0, s)\over { \ ImF(q=0, s)}}
\end{equation}
Multiple-diffraction models differ one from another by the
particular choice of parameterizations made for $G_{A}$ and
$G_{B}$ and the elementary amplitude $f$. In the case of identical
particles, as is our case, $G_{A}=G_{B}=G$. For our purposes we
follow the parametrization developed in a highly outstanding  work
\cite{MM1} which has the advantage of using a small set of free
parameters, five in total, which are in principle energy
dependent: two of them $(\alpha^{2}, \beta^{2})$ associated with
the form factor
\begin{equation}
   G = (1+ \frac{q^{2}}{\alpha^{2}})^{-1} (1+ \frac{q^{2}}{\beta^{2}})^{-1}
\end{equation}
and the other three $(C, a, \lambda )$ associated with the elementary complex amplitude $f$
\begin{equation}
 f(q,s) = Ref(q,s)+ iImf(q,s)
\end{equation}
where
\begin{equation}
Imf(q,s)= C \frac{1-\frac{q^{2}}{a^{2}}}{1-\frac{q^{4}}{a^{4}}} \
\end{equation}
and
\begin{equation}
Re f(q,s) = \lambda(s)Imf(q,s) \
\end{equation}
so that,
\begin{eqnarray}
ImF(q=0,s) =  \ \ \ \ \ \ \ \ \ \ \ \ \ \ \ \ \ \ \ \ \ \ \ \ \ \ \ \ \ \ \ \ \ \ \ \ \ \  &  & \nonumber \\
 \int_{0}^{\infty} \!\! \left[ 1 - e^{- \Omega(b,s)}  \cos \left\{ \lambda \Omega(b,s) \right\} \right]  b \ db \ J_{o}(q,b) \mid_{q=0}& &
\end{eqnarray}
with the opacity $\Omega(b,s)$  given as:
\begin{equation} \Omega(b,s) = \int_{o}^{\infty} G^{2} \ Imf(q,s)\ J_{o}(q,b)
                                \ qdq
\end{equation}
which explicit expression is:
\begin{eqnarray}
\Omega(b,s) = C \{ E_{1}K_{0}(\alpha b) + E_{2}K_{0}(\beta b) + E_{3}K_{ei}(ab) + & \nonumber \\
                            E_{4}K_{er}(ab) + b \left[ E_{5}K_{1} (\alpha b) + E_{6}K_{1} (\beta b) \right] \} \ &
\end{eqnarray}
where   $k_{0}, k_{1}, k_{ei}, $  and $ k_{er}$ are the modified Bessel  functions, and $E_{1}$  to $E_{6}$  are functions of the
five free parameters. The proton-proton total cross-section is directly determined by the expression
\begin{equation}
\sigma_{tot}^{pp} = 4\pi \int_{o}^{\infty} \!\! b \, db  \left\{ 1 - e^{- \Omega(b,s)}
 \cos \left[ \lambda \Omega(b,s) \right] \right\} J_{o}(q,b) \mid_{q=0}
\end{equation}
expression that  was numerically solved in \cite{PP2}, \cite{PP1}.
It should be noted that, according to the principle of
Analyticity, the scattering forward amplitude for
particle-particle and particle-antiparticle appears from the same
analytical function \cite{Block1}. The behavior of total
cross-sections of both reactions is assumed to follow one of the
following ways: up to the ISR energies the differences are
attributed to Regge contributions, which are expected to disappear
at higher energies
 \cite{DL1}, \cite{BSW}, or on the contrary, differences are interpreted
in terms of the Maximal Odderon Hypothesis, which predicts that
they increase as the energy overpass the highest energy of the
ISR. At this level let us now  emphasize an  essential point in
all which  follows. First, we  quote the  success in the
prediction of  the value  of $\sigma_{tot}^{\bar{p}p}$  at the
$S\bar{p}pS$  Collider  \cite{UA41} from the ISR data (mainly  pp)
using   expressions  (1) and (2), where  $\sigma_{tot}^{pp}$ and
$\sigma_{tot}^{\bar{p}p}$ were  taken  asymptotically  equal
\cite{Amaldi2}. Secondly, according to \cite{CAR1} the difference
$\Delta\sigma$ between $\sigma_{tot}^{\bar{p}p}$  and
$\sigma_{tot}^{pp}$, $\Delta\sigma = \sigma_{tot}^{\bar{p}p} -
\sigma_{tot}^{pp}$, up to energies  $\leq$ 2000 GeV in the
laboratory  ($\sim$ 60 GeV in the center of mass) tends toward
zero as $s^{-0.56}$, Figure 2.  And  finally, even if  it is
argued that $\sigma_{tot}^{pp}$  and $\sigma_{tot}^{\bar{p}p}$ are
anyway different for higher energies, we  have  indicated in
Section I that  current  evidence   points the other way  round.
Taking this triple line of  evidence into account, in our
multiple-diffraction analysis it is assumed the same behavior for
$\sigma_{tot}^{pp}$ and $\sigma_{tot}^{\bar{p}p}$ at high energy.
So, here after $\sigma_{tot}^{pp} = \sigma_{tot}^{\bar{p}p} =
\sigma_{tot}$. It is noteworthy that some parametrization models,
as the RRP  \cite{Cude}  predict the same value for both cross
sections at $\sqrt{s} > 70$ GeV and the same value for the
corresponding $\rho$  at $\sqrt{s} > 110$ GeV.

\begin{figure}
\centering
\includegraphics[width=11cm]{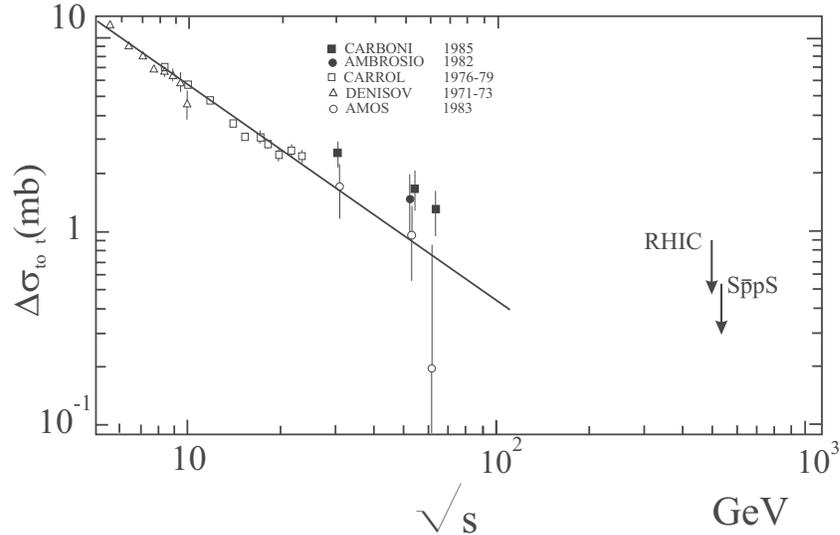}
\caption{Plot of $\Delta \sigma_{tot} \equiv
\sigma^{tot}_{\bar{p}p}-\sigma^{tot}_{pp} \sim s^{-0.56}$ indicating that
$\Delta \sigma^{tot}$  tends towards zero up to energies $\sim$ 60 GeV in the
center of mass.  Data points  from \cite{CAR1}}
\end{figure}

\subsection{ Evaluation of the Model Parameters}

For the evaluation of the free parameters it was assumed,
following \cite{MM1} that two of them are constants: $a^{2}$ =
$8.2 \ GeV ^{2}$ and $\beta^{2}$ = $1.8 \ GeV^{2}$. The other
three parameters are: $C(s)$ and $\alpha^{-2}(s)$ which determine
the imaginary part of the hadronic amplitude [Eq. (11)] and
consequently the total cross section [Eq. (6)] ,  and on the other
hand,  $\lambda(s)$, which controls the real part of the amplitude
[Eq. (12)]. Setting $\lambda(s)$ = $0$ makes the amplitude purely
imaginary, so that a zero (a minimum) is produced in the dip
region, where only the real part of the amplitude becomes
important \cite{Menon}.  We fit  experimental data (Fig. 3) of the
elastic  differential  cross  section, ${d\sigma\over{dq^{2}}}$,
 with Eq. (7), choosing  the set ($C, \alpha^{-2}$) for which the theoretical
and the experimental central values are equated at the precise
$-q^{2}$ value where  the data show its  first minimum (the
``dip'' position), which  coincides with the minimum of  the
imaginary part of the  elastic amplitude. Data of  $pp$
differential cross section at $13.8$ and $19.4$ GeV  was taken
from \cite{Rubi},  for $23.5 - 62.5$  GeV from \cite{Schu}, for
$546$ GeV  $(\bar{p}p)$  from \cite{UA4-1}, (for ${\it -t} < 0.5$
(GeV/c)$^{2}$) and \cite{UA4-2} (for $0.5 \leq {\it -t} \leq 1.55$
(GeVC)$^{2}$), and for $1800$ GeV from \cite{Amos-1}(for $0.034
\leq {\it -t} \leq 0.65$ (GeVC)$^{2}$). The procedure is carried
out for each one of the energies where there is available
accelerator data of ${d\sigma\over{dq^{2}}}$ in the interval
$13.8 \leq \sqrt{s} \leq 1800$ GeV, as illustrated in  Fig 4 and
Fig 5. Because  error bars of data at the dip position are small
the fitting procedure is based on the central values. It must be
emphasized that, it is precisely because the minimum of the
imaginary amplitude is produced in the dip region, that  data at
$1800$ GeV is quite suitable  to our procedure, since the
predicted minimum falls at  $t = 0.585$ $GeV^{2}$ where there is
available data (${d\sigma\over{dt}} = 0.0101$). The next data
value at $t = 0.627$ $GeV^{2}$ shows an slight increase.
Furthermore, the shift of the dip region toward lower values of
$t$ as the considered energy increases is qualitatively consistent
with the expectation relative to the shoulder at $546$ GeV ($t =
0.9$ $GeV^{2}$). Once the values of $C(s)$ and $\alpha^{-2}(s)$
are determined for  each  energy they are introduced, together
with the constant parameters $a^{2}$ and $\beta^{2}$ into Eq. (8).
Giving values to the parameter $\lambda(s)$ up to obtain the
experimental value of $\rho(s)$,  we determine the $\lambda$ value
for each energy where there is available accelerator data in the
same interval  $13.8 \leq \sqrt{s} \leq 1800$ GeV.  The obtained
central values of the three energy-dependent free parameters,
$C(s)$,  $\alpha^{-2}(s)$and  $\lambda(s)$ are listed in Table 2.
Therefore, the employed method to evaluate the model parameters
only requires  data of  $(d\sigma\over{dt})$ and $\rho(s)$.
\begin{table}
\centering
\caption{Values of the  parameters $C$, $\alpha^{-2}$,
$\lambda$ at each energy. They are obtained by equating the
accelerator data and the model prediction for the elastic
differential cross-sections and for the parameter $\rho$ in the
interval $13 \le \sqrt{s} \le 62.5$ and $ 546 \le \sqrt{s} \le
1800$ GeV. } \vspace{0.5cm}
\begin{tabular}{rccc}
\hline \hline
$\sqrt{s} $ \ \ &  $ C(s) $ & $\alpha^{-2}(s) $ & $\lambda(s) $ \\
(GeV) & (GeV$^{-2}$) & (GeV$^{-2}$) & \\
\hline
13.8   &   9.97   &  2.092    & -0.126   \\
19.4   &  10.05   &  2.128    & -0.043   \\
23.5   &  10.25   &  2.174    &  0.025   \\
30.7   &  10.37   &  2.222    &  0.053   \\
44.7   &  10.89   &  2.299    &  0.079   \\
52.8   &  11.15   &  2.350    &  0.099   \\
62.5   &  11.42   &  2.380    &  0.115   \\
546    &  16.90   &  3.060    &  0.182    \\
1800   &  21.52   &  3.570    &  0.194     \\
\hline \hline
\end{tabular}
\end{table}

\begin{figure}
\centering
\includegraphics[width=11.0cm]{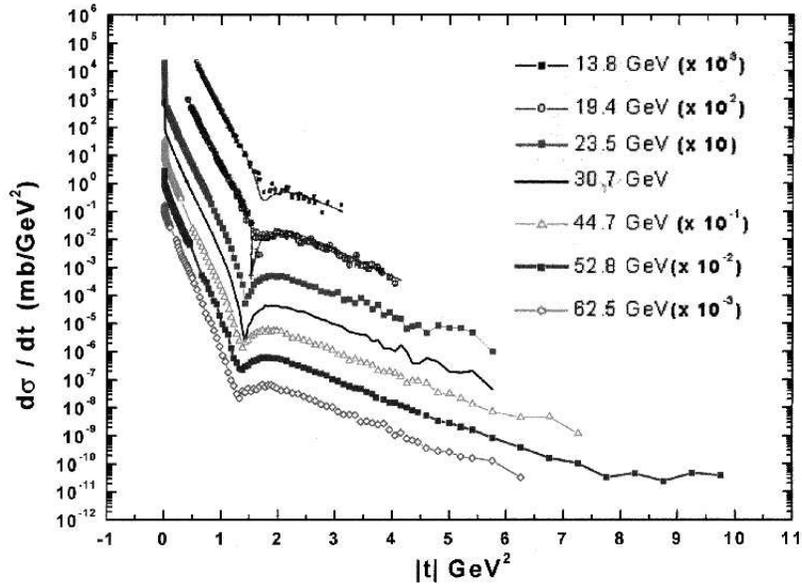}
\caption{Data of $pp$ differential cross section at $\sqrt{s} \le 62.5$
according to \cite{Rubi},\cite{Schu}, \cite{UA4-1}, \cite{UA4-2}
and \cite{Amos-1}.}
\end{figure}

\begin{figure}
\centering
\includegraphics[width=9.0cm]{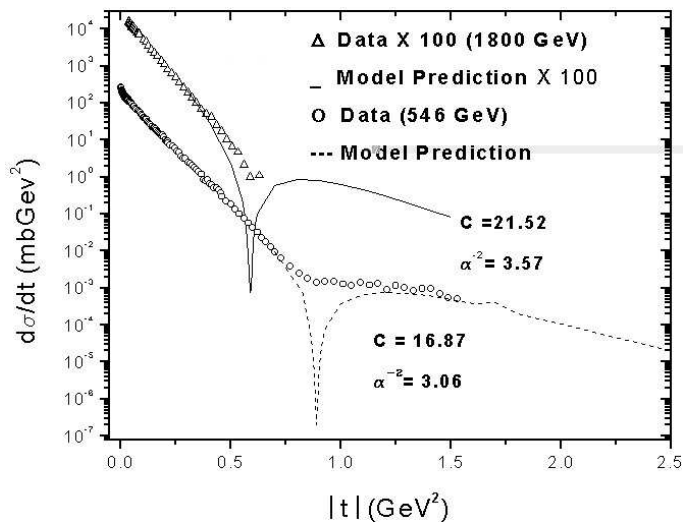}
\caption{Fits of the predicted imaginary amplitude to data of the
$\bar{p}p$ differential cross sections to determine $C(s)$ and
$\alpha^{-2}(s)$ by equating experimental and theoretical values
at the specific $\mid t \mid$ of the ``dip'' position for
$\sqrt{s} = 546 GeV$ and $\sqrt{s} =1800 GeV$.}
\end{figure}

\begin{figure}
\centering
\includegraphics[width=9.0cm]{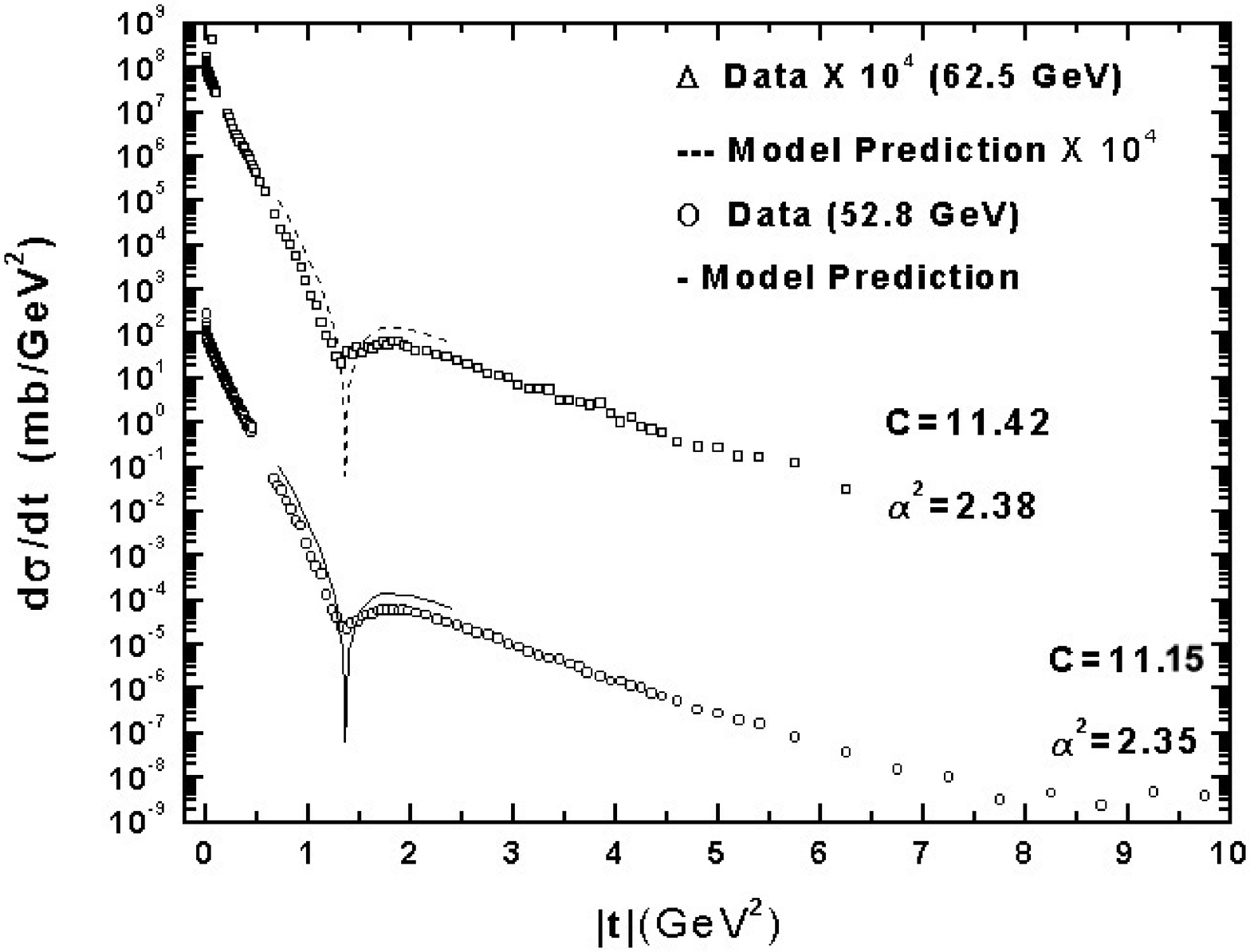}
\caption{Idem that Fig. 4, for $\sqrt{s} = 52.8$ and
$\sqrt{s} = 62.5 GeV$.}
\end{figure}

\section{PARAMETRIZATION FOR EXTRAPOLATION TO HIGH ENERGIES}
Concerning the physics behind a parametrization and fitting
procedures, it should be emphasized  that there is not yet an
answer on microscopic basis, i.e., on terms of the QCD theory,
because diffractive physics is essentially a non-perturvative
phenomenon and confinement is the unsolved problem of QCD.
However, on basis to the present model  it can be said that the
closeness of our fits to the colliders data   (ISR, $S\bar{p}pS$
and Tevatron) gives us an indication that the basic tenets of the
model are right, as was described for instance in reference [32].
In spite of the approximation made in ignoring crossing symmetry,
the parametrization carried out here is not an arbitrary one as
will be described below.

On this basis, with the aim of interpolating values at energies where there is no available accelerator data and for
extrapolation to higher energies, we proceeded to perform
a parametrization of the three energy-dependent free parameters.
According to the obtained values of those parameters, as described
in Section III-A  (Table 2) and following the procedure to
be described in Section V, a second-order fitting  of the values
of $C(s)$ and $\alpha^{-2}(s)$ and a exponential fitting
of $\lambda(s)$ has been obtained from the following analytical expressions:
\begin{equation}
C(s) = 19.24521 - 2.86114\ ln(s)+0.22616\ ln(s)^{2}
\end{equation}
\begin{equation}
\alpha^{-2}(s) =  1.8956 - 0.03937\ ln(s)+0.01301\ ln(s)^{2}
\end{equation}
\begin{eqnarray}
\lambda(s) = 0.01686 + 0.00125 \left( 1 - \ e^{-ln(s/400)/0.18549} \right) +
             & \nonumber \\
0.19775 \left( 1 - \ e^{-ln(s/400)/3.74642} \right) \ &
\end{eqnarray}
Results are displayed in Table 2  and  illustrated in Figures (6) -(8)
as the central solid lines. \\
However, it must be taken into account that the reliability of the
functional parameterizations for extrapolations must have a
physical support, since it is clear that several parameterizations
that may describe correctly the set of data in the experimental
range $13.8 \leq \sqrt{s} \leq 1800$ GeV, may not be consistent
any more, but differ substantially when extrapolated to high
energies. Therefore, some restrictions must be imposed in the
selection of parameterizations according to the physical
information available.  As we mentioned before, our fittings of
$C(s)$ and  $\alpha^{-2}(s)$ in the limited experimental range
were based on experimental data of the differential cross-section
and the $\rho(s)$ parameter, obtaining values that increase with
energy (Table 2) with positive concavity (Figs. 6-7).
Experimentally, total cross-sections increase with energy like
$\ln (s)$ or $\ ln^{2}(s)$ in the concerned energy range, and on
the other hand, soft processes are expected to have a $\ln (s)$
behavior: from the optical theorem, the interdependence between
the free parameters and the physical observable (Eq. 6) may be
connected with the unitary condition, for which lowest order
cross-sections  within the frame of gauge field theories  present
$\ln (s)$  terms \cite{ChenWu}. Based on our fittings and
extrapolation results and on the previously mentioned constraints,
the presence of  $\ln (s)$  terms in the two-energy dependent
parameters $C(s)$ and  $\alpha^{-2}(s)$ is a natural consequence,
and therefore the hypothesis of polynomial functions of $\ln(s)$
seems a reasonable one.

Concerning the parameter $\lambda(s)$ let us pointed out that
a basic property of the Glauber Multiple
Diffraction Model is to associate elastic scattering cross-sections
of nucleons with the scattering amplitude
of their composite partons \cite{GV1}: within this framework,
the proportionality between real and imaginary
parts of the parton-parton amplitudes (Eq. 12),
$\lambda(s)= Re f(q,s)/Imf(q,s)$ implies that $\lambda(s)$
at the partonic scale behaves similarly as $\rho(s)$ does at
the hadronic scale. The influence of $\lambda(s)$
on the hadronic amplitude has been empirically analyzed in \cite{MM1},
showing that if $\lambda(s)$ increases
 (or decreases) also $\rho(s)$ increases (or decreases) (Fig. 9),
and $\lambda(s) = 0$ at the same energy value where
$\rho(s) = 0$. However, due to the lack of $\rho(s)$ data above
the experimental energy range used in this work,
the parametrization  of $\lambda(s)$ at high energies is based on
the conventional assumption that beyond
$\sqrt{s} \sim  100$ GeV, $\rho(s)$ presents a maximum and then goes
asymptotically to zero \cite{GM2},
the rate of  convergence depending  on  model particularities.
Based on this consideration and on the empirical behavior
of $\lambda(s)$, shown in Table 2, we propose the parametrization
given in Eq. (19),  where $S_{o} = 400 $ GeV$^{2}$ is the
value at which $\lambda(s)$ converges to zero, and the numerical
coefficients control the maximum and asymptotic behavior.

Let us now remind that blackening and expansion are very well known
properties of elastic scattering. Within
the context of the present empirical analysis,  blackening and expansion
are related to the elementary parton-parton amplitude and
the hadronic form factors through the energy-dependent parameters $C(s)$
and $\alpha(s)$  respectively.  Since in the very forward direction
the scattering amplitude is basically of diffractive nature, the
eikonal becomes a purely imaginary one, then  for hadron-hadron
$\xi (b,s) = C(s)\left< G^{2} Im f(q,s) \right> = Im \Omega(b,s)$,
so  that in terms of Eqs. (9) and (11) the opacity
satisfies $\Omega(b,s)  \geq 1$, and the free parameter $C(s)$ behaves
as an absorption factor (optical theorem).  On the other hand,
the free parameter $\alpha(s)^{2}$ may be connected
to the hadronic radius \cite{MM1} as $R^{2}(s) = - 6 [dG(q,s)/ dt]_{t = 0}$.
Then, from Eq. (9)  \[ R^{2}(s) = 0.2332\left [\frac{1}{ \alpha^{2}(s)} + \frac{1}{\beta^{2}(s)}\right]        (fm)^{2}.\]

Therefore, from Eq. (18) and the adopted value $\beta^{2} = 1.8 $
GeV$^{2}$ the radius is an increasing function of energy  and such
a behavior is connected to the expansion effect. The result is
that hadrons become blacker and larger with energy increase,
consistently with the so called ``Bell Effect'' \cite{HenziVali}.
Since the hadronic scattering amplitude is purely imaginary the
free parameters may be associated with the physical observable by
means of Eq. (6) - (7).

\section{THE EXTRAPOLATION PROCEDURE}
The procedure followed to arrive to the predictions of $\sigma_{tot}^{pp}$
at high energies with statistic confidence intervals based on
the {\em {\em {\em forecasting}}} method described in Appendix A is  the next:  -(i) using the values displayed in Table 2,  we established  prediction  equations of
the type of eq. (A4) within the data range and  eq.  (A6) )   out of the range,  for each of the energy-dependent parameters:
(ii)  using the least squares method in matrix formalism, as described in the Appendix (eqs. A10-A12) we obtained
the constants $\hat{\beta}_{0}, \hat{\beta}_{1}, \hat{\beta}_{2}$ for each one of the parameters. The autocorrelation
constant was determined as described in \cite{Full,Chat},
(iii)   with the obtained values the central values of the three parameters were derived (eqs. 17-19), leading to a second-order
fitting  of the values of $C(s)$ and $\alpha^{-2}(s)$ and a exponential fitting of $\lambda(s)$; results are shown in Table 2,
(iv)  we evaluated the variance for each one of the new forecasted value (eq. A14) ,
(v) by means of eqs. (A15)-(A16) we estimated the confidence intervals for each one of the interpolated-extrapolated
values, such that by a fitting of the extreme values of these confidence intervals we have  built the error bands of each one
of the parameters, what is shown in Figs. (6)-(8),
(vi) the {\it Central values} of $\sigma_{tot}^{pp}$ for each point are obtained by means of the introduction in eqs. (15)-(16) of
the values of $C(s)$, $\alpha^{-2}(s)$ and $\lambda(s)$ (displayed in Table 2),
(vii) finally, the  overall confidence  band for the predicted  $\sigma_{tot}^{pp}$  is obtained, not from eqs. (A15)-(A16),
but from the substitution of the extreme values of the error bands of the three energy-dependent parameters into eqs.
(15)-(16), followed by the corresponding fittings (Figs. 10-11).
\begin{figure}
\centering
\includegraphics[width=8.0cm]{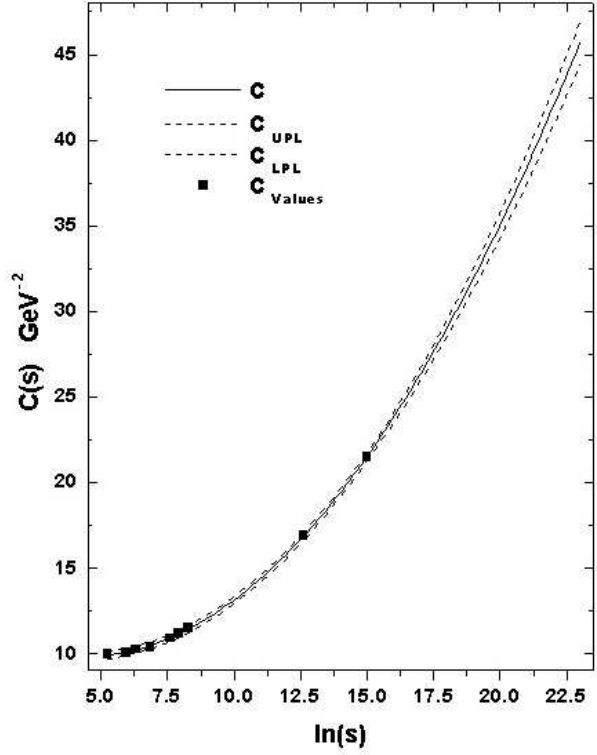} \caption{The
parameter $C(s)$ from Table 2 with its confidence interval.}
\end{figure}
\begin{figure}
\centering
\includegraphics[width=8.0cm]{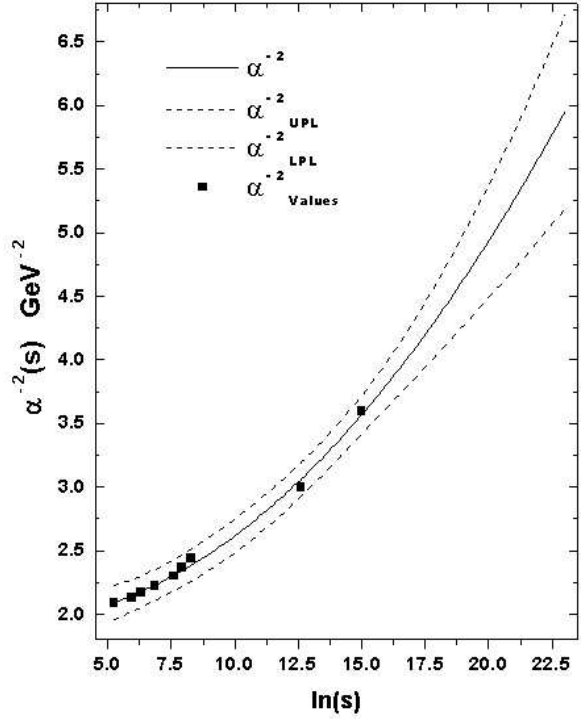}
\caption{The parameter $\alpha^{-2}(s)$ from Table 2 with its confidence interval.}
\end{figure}
\begin{figure}
\centering
\includegraphics[width=8.0cm]{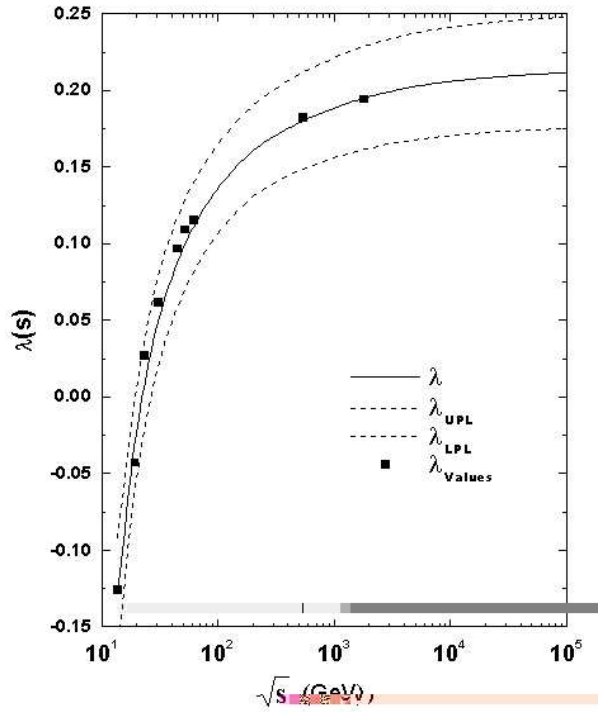}
\caption{The parameter $\lambda(s)$ from Table 2 with its confidence interval.}
\end{figure}
\begin{figure}
\centering
\includegraphics[width=7.0cm]{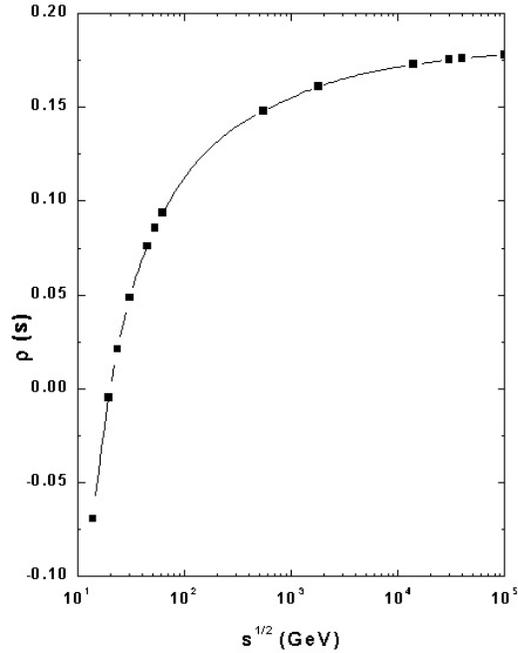}
\caption{Data behavior of the parameter $\rho(s)$ at $\sqrt{s} \le 62.5$ }
\end{figure}

\begin{table}
\centering
\caption{Fitted  values (interpolation and
extrapolation) of $C(s)$, $\alpha^{-2}(s)$ and $\lambda (s)$ .}
\vspace{0.5cm}
\begin{tabular}{rccc}
\hline \hline
$\sqrt s$ \ \ & $C(s)$ & $\alpha^{-2}(s)$ & $\lambda(s)$ \\
(GeV) & (GeV$^{-2}$) & (GeV$^{-2}$) & \\
\hline
13.8 & 9.9039  & 2.0945 & -0.12816  \\
19.4 & 10.082 & 2.1469 & -0.02848   \\
23.5 & 10.225 & 2.1798 & 0.00975     \\
30.7 & 10.474 & 2.2296 & 0.04942     \\
44.7 & 10.923 & 2.3075 & 0.08786     \\
52.8 & 11.159 & 2.3451 & 0.10064      \\
62.5 & 11.421 & 2.3849 & 0.11172      \\
546  & 16.872  & 3.0634 & 0.18035     \\
1800 & 21.518  & 3.5685 & 0.19501   \\
14000 & 32.239  & 4.6555 & 0.20703   \\
16000 & 33.056  & 4.7359 & 0.20749   \\
30000 & 37.102  & 5.1298 & 0.20927   \\
40000 & 39.062  & 5.3188 & 0.20993   \\
100000 & 45.757  & 5.9568 & 0.21153   \\
\hline \hline
\end{tabular}
\end{table}

\section{RESULTS}
Data of the total cross sections with their respective errors
are summarized in Table IV.
At $ \sqrt{s} \leq 62.5$ GeV data was taken from \cite{Schu},
for $546$ GeV from \cite{UA41}. For the value at $1800$ GeV there exist
three different measurements: the value of the
{\it CDF collaboration} $(80.3 \pm 2.24 mb)$
 \cite{CDF}, the value of the  {\it E710 collaboration}
$(72.8 \pm 3.1 mb)$  \cite{E710} and the value of
the  {\it E811 collaboration} $(71.7 \pm 2.02 mb)$ \cite{Avil}.
There is a high controversy on whether the correct value is that of
the  {\it CDF collaboration}   or those lower values of the
{\it E710 collaboration} and  the {\it E811 collaboration}, mainly related  to
the  estimation of the  different backgrounds.
For  instance the CDF Collaboration has  decided  to  use  only their
$\sigma_{tot}^{pp}$ value  to quote luminosity  values  for    all of  their
physics  program  \cite{Albrow} whereas the other
Tevatron Collaboration, {\it D0},  has adopted   the  average of the three measurements \cite{Albrow}.
 For the goal of this work, one of us (J.V.) suggested to follow  this  second
option and use  the arithmetic weighted mean  of  the three
values, ($74.91 \pm 1.35$ mb). The consequence in our results of
the employment of  the values of those  three collaborations is
mentioned in the next section.

We  quote  in Table IV the values
of  the  predicted  $\sigma_{tot}^{pp}$
for  two different  energy intervals:  $\sigma_{tot}^{1800}$,
($ 13.8 \leq \sqrt{s} \leq 1800$  GeV), and   $\sigma_{tot}^{546}$,
($ 13.8 \leq \sqrt{s} \leq 546$ GeV).
Figure 10  represents graphically the first case together with cosmic ray data. For further  discussion,
we include in Table IV predicted values from  two standard  accelerator-based  data
extrapolations: the  first one, $\sigma^{[54]}_{546}$ makes  a fit  to  all $\sigma_{tot}^{pp}$
and  $\sigma_{tot}^{\bar{p}p}$ data in the interval $10 \leq \sqrt{s} \leq 546$ GeV using expression (2)  \cite{BV},
and the second one, $\sigma^{[5]}_{\rho}$ (Fig. 1), in the same  energy range, makes the simultaneous  fit  to all
$\sigma_{tot}^{pp}$, $\sigma_{tot}^{\bar{p}p}$, $\rho_{tot}^{pp}$  and $\rho_{tot}^{\bar{p}p}$
data  through the  method described  in Section I, \cite{Augier1} .

\begin{table}
\caption{Central, upper and lower values for $\sigma_{tot}$
obtained with the {\sl {\em {\em forecasting}}} method. The
$\sigma^{1800}_{tot}$, $\sigma^{546}_{tot}$, $\sigma^{62.5}_{tot}$
columns includes data up to  1800, 546, 62.5 GeV respectively. The
quoted $\sigma_{546}^{[54]}$ and $\sigma_{\rho}^{[5]}$ values are
from \cite{BV}, and\cite{Augier1} respectively.}
\vspace{0.5cm}
\begin{tabular}{rcccccc}
\hline\hline
$\sqrt s$ (GeV) &  $\sigma^{exp}_ {tot}$ (mb)& $\sigma^{1800}_{tot}$ (mb)
& $\sigma_{546}^{[54]}$ (mb)& $\sigma^{546}_{tot} $ (mb)& $\sigma_{\rho}^{[5]}$ (mb)& $\sigma^{62.5}_{tot} $ (mb) \\
 \hline
13.8        & 38.36  $\pm$ 0.04 & $38.29_{-0.98}^{+1.06}$
      & - - - &$38.30_{-0.92}^{+0.94}$  &- - -& $38.51_{-0.56}^{+0.62}$    \\
19.4        & 38.97 $\pm $ 0.04 & $38.92_{-0.97}^{+0.98}$
      & - - - &$38.82_{-0.87}^{+0.88}$   &- - -&$38.82_{-0.54}^{+0.59}$    \\
23.5        & 38.94  $\pm$ 0.17 & $39.44_{-0.96}^{+0.97}$
      & - - - &$39.44_{-0.86}^{+0.88}$    &- -  -&$39.26_{-0.58}^{+0.62}$    \\
30.7        & 40.14 $\pm $ 0.17 & $40.34_{-0.95}^{+0.99}$
      & - - - &$40.37_{-0.88}^{+0.88}$   &- - -&$40.15_{-0.61}^{+0.66}$      \\
44.7        & 41.79 $\pm $ 0.16 & $41.93_{-0.96}^{+1.06}$
      & - - - &$42.00_{-0.92}^{+0.93}$    &- - -&$41.93_{-0.62}^{+0.66}$     \\
52.8        & 42.67 $\pm$ 0.19 & $42.76_{-0.98}^{+1.09}$
      & - - - &$42.84_{-0.94}^{+0.95}$    &- - -&$42.93_{-0.64}^{+0.69}$     \\
62.5        & 43.32 $\pm$ 0.23 & $43.67_{-0.99}^{+1.13}$
      & - - - &$43.77_{-0.97}^{+0.98}$    &- - -&$44.05_{-0.72}^{+0.78}$       \\
546         & 61.5 $\pm$ 1.5 & $61.62_{-0.94}^{+1.58}$
      & - - - &$61.78_{-1.29}^{+1.33}$  &- - -&$69.39_{-7.4}^{+8.4}$      \\
1800       & 74.91 $\pm  $1.35$^{(*)}$ & $76.17_{-1.07}^{+2.02}$
   & 76.7 $\pm$ 4.0 & $76.00_{-2.34}^{+2.41}$ & 76.5 $\pm$ 2.3&$91.74_{-14.7}^{+16.9}$    \\
14000     & - - - & $108.27_{-3.17}^{+4.72}$ & 112 $\pm$ 13.0
     & $106.5_{-6.55}^{+6.56}$ &- - -&$143.86_{-33.5}^{+38.6}$    \\
16000     & - - - & $110.67_{-3.40}^{+5.00}$ & - - -
     & $108.7_{-6.94}^{+6.94}$ & 111.0 $\pm$ 8.0&$147.85_{-35.1}^{+40.3}$  \\
30000     & - - - & $122.41_{-4.62}^{+6.40}$ & - - -
     & $119.6_{-8.96}^{+8.83}$  &- - -&$167.64_{-42.6}^{+48.9}$   \\
40000     & - - - & $128.05_{-5.27}^{+7.08}$ & - - - &
   $124.7_{-9.97}^{+9.83}$  & 130.0 $\pm$ 13.0&$177.23_{-46.3}^{+53.1}$   \\
100000   & - - - & $147.14_{-7.68}^{+9.63}$ & - - - &
   $142.0_{-13.7}^{+13.33}$  &- - -&$210.06_{-59.1}^{+67.6}$   \\ [0.5ex]
\hline \hline
\end{tabular}
(*) This value is the weighted arithmetic mean of the E710 ($72.8\pm3.1$ mb),
CDF ($80.3\pm2.3$ mb) and  E811 ($71.7 \pm 2.0$) values.
\end{table}

\section{DISCUSSION}
To start  with, let first  examine  what  happens  when our main assumption,
the  asymptotic equality of $\sigma_{tot}^{pp}$
and $\sigma_{tot}^{\bar{p}p}$, is  not used.
Analysis of Figure 11 shows that if we limit our fitting calculations
to the accelerator domain $\sqrt{s} \leq 62.5$ GeV, where
$\sigma_{tot}^{pp}$  data exist, the extrapolation at cosmic  ray energies
produces an error band so large
that practically any cosmic ray result become compatible with results
at accelerator energies.  It can be seen that in this case
the $\sigma_{tot}^{pp}$ values obtained when extrapolated to ultra
high energies seem to confirm the highest quoted values
of the cosmic ray experiments \cite{Niko,GSY}.
Also it can be noted that such extrapolation  to ultra high energies
may claim not only agreement with the analysis carried out
in \cite{Niko} and the experimental data of the Fly's Eye \cite{GSY},
but even with the Akeno collaboration \cite{Akeno1},  because
their experimental  errors are so big that they overlap with the errors
reported in \cite{GSY}, and of course falling within the predicted
error band for that case $\sqrt{s} \leq 62.5 $ GeV (Fig. 11).
If that  were true, it  would  imply the extrapolations cherished
by experimentalists are wrong.   But  the prediction shown in Fig. 11
gives  $\sigma_{tot}^{pp} =  $  69  mb at the CERN  $S\bar{p}pS$ Collider
($546$ GeV), and $91.6 mb $ at the Fermilab Tevatron ($1.8$ TeV).
Comparing with  Table I we see that the measured $\sigma_{tot}^{\bar{p}p}$
at $546$  GeV is  smaller than the predicted $\sigma_{tot}^{pp}$ by
near $8$ mb, and in the case of $1.8$ TeV by more than $15$ mb,
which no available model is able to explain \cite{Blois99}.
However when our  main  hypothesis,
$\sigma_{tot}^{\bar{p}p} = \sigma_{tot}^{pp}$ asymptotically  is  used,  then
the existing  $\sigma_{tot}^{\bar{p}p}$ data at higher accelerator
energies may safely be  included.  This  allows  to enlarge by a
great amount the lever  arm for the  extrapolation  and  both
the predicted values and the error band change considerably.
This can be clearly seen in Fig. 10, as well as in Table  IV, where
we have added the available $\sigma_{tot}^{\bar{p}p}$
up  to  0.546 TeV  (the $\sigma^{546}_{tot}$ column)  and
up  to 1.8 TeV  (the $\sigma^{1800}_{tot}$ column).  Now
the predicted value of $\sigma_{tot}^{pp}$ from
our extrapolation $\sigma^{1800}_{tot}$, for instance
at $\sqrt{s} = 40 $ TeV, $\sigma_{tot}^{pp} =
128.0_{-5.27}^{+7.08}$ mb,  is seen incompatible with those
in \cite{Niko,GSY} by several standard deviations, though no so
different to the Fly's Eye  or Akeno results
and the predicted value  in \cite{Augier1}.

\begin{figure}
\centering
\includegraphics[width=10cm]{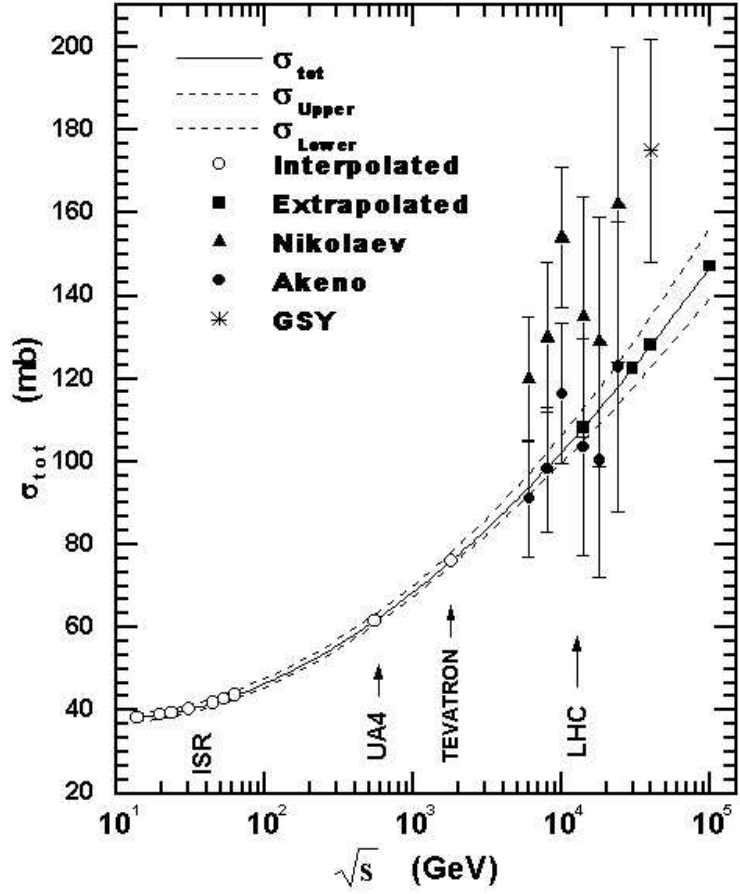}
\caption{Prediction of the total cross-section
using data up to 1800 GeV (solid line), $\sigma^{1800}_{tot}$. The error band
(dashed lines) is obtained by the {\sl {\em {\em forecasting}}} method.}
\end{figure}

Concerning the quoted  error bands, in sections V,  it has been
shown in  the Appendix that due to the inclusion of {\it Residual
Correlations} in the {\em Forecasting} method, this is always a
high precise technique whatever the employed parametrization
model, but this systematic advantage occurs, provided we are
dealing with the same parametrization model under the same input
conditions. Therefore, it must be emphasized that  it is not valid
a comparison between different statistical techniques on basis to
different parametrization models,  or running of  the same
parametrization  with different input values. However, in spite of
this,  we would like to make a consideration of strictly
qualitative nature, in the sense that other parameterizations
\cite{Augier1}, \cite{BV} with similar input quantities (for
instance  $\rho$ and ${d\sigma\over{dt}}$) lead to central values
that are only slightly higher than ours, whereas the quoted errors
are larger than ours by nearly a factor of 3, as can be seen in
Table 4, or  in Fig. 4 and Fig.10.  For  instance, the  quoted
error in the  fit  $\sigma_{tot}^{1800}$ at  100 TeV  in the
center of mass (which corresponds to $\simeq 10^{19}$ eV  in the
lab), $147_{-7.68}^{+9.63}$, is comparable (or even better) to the
error obtained at much more lower energies in other works as can
be seen in Table IV, at  $16$ TeV and $14$ TeV ($ \simeq 10^{17}$
eV in the lab),  corresponding to the energy range of future LHC
CERN Collider. Another parametrization which predictions of the
cental values are not so different of  ours, is that of the {\sl
Ensemble A}, of the work \cite{Avila1, Caravalho1}. Obviously, the
previous analysis is only of qualitative nature, but it is however
quite important because such a discussion opens new
interrogatives: - though, it is clear that from  statistic theory
the {\em {\em forecasting}} technique is a highly  precise method
(section V, Appendix and references therein) , its advantage  is,
however, only a second order effect when applied to the same
conditions (as shown in Eqs. 20 and 27),  why this higher
preciseness  seems to be amplified
 when qualitative comparisons are made with a different parametrization model the same  different inputs?, is it because the employed parametrization
model is better?. It is clear that the answer must be based on  a deep quantitative  analysis,  by applying the
{\em {\em forecasting}} method  to different parametrization models of total cross-section,
 in order to discern which model is more confident, and from there to draw more  reliable physical inferences.
 Such an analysis is out of the scope of the present paper, and will be the subject of a forthcoming work.

\begin{figure}
\centering
\includegraphics[width=8.5cm]{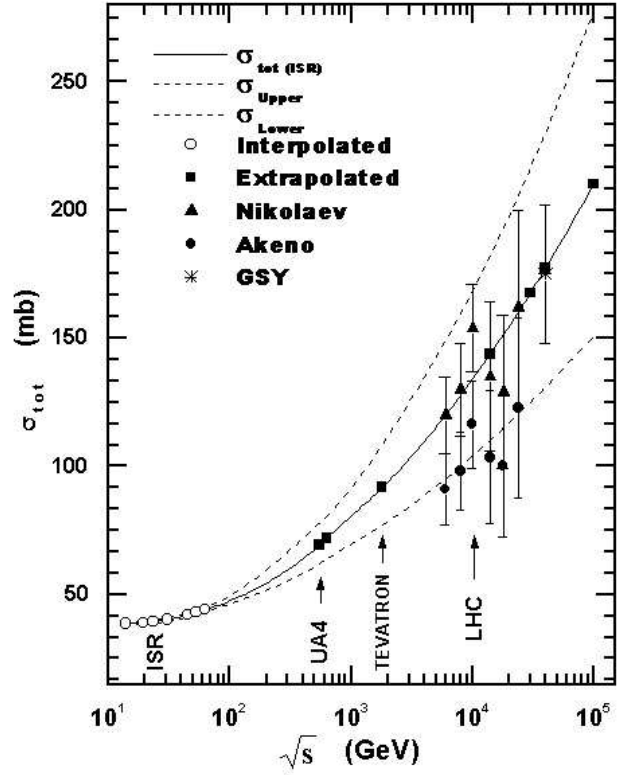}
\caption{Prediction of the total cross-section
using $pp$ data up to 62.5 GeV (solid line). The error band
(dashed lines) is obtained by the {\sl {\em {\em forecasting}}} method.}
\end{figure}\

\section{CONCLUSIONS}
It has been shown in this work that highly confident predictions
of high energy $\sigma_{tot}^{pp}$ values are strongly dependent
on the energy range covered by experimental data and the available
number of those data values. In particular, if we limit our study
of determining $\sigma_{tot}^{pp}$ at cosmic ray energies from
extrapolation of $\sigma_{tot}^{pp}$ accelerator data in  the
range $\sqrt s \leq  62.5$ GeV, then the obtained results are
compatible with most of cosmic ray experiments and other
prediction models, because the predicted error band is so wide
that covers their corresponding quoted errors  (Fig. 11). However,
as the included data in our calculations extends to higher
energies, including $\sigma_{tot}^{\bar{p}p}$ data  up to  1.8
TeV, that is, when all experimental available data  is taken into
account, the range of possibilities  is much reduced. It should be
noted that our predictions are compatibles with other prediction
studies \cite{Augier1}. Taken all this  results at face value,
i.e. as indicating the most probable $\sigma_{tot}^{pp}$ value, we
conclude that if predictions from accelerator data are correct,
hence it should be of great help to normalize the corresponding
values from cosmic ray experiments, as for instance by keeping the
($k$) parameter as a free one. The $k$ value found will greatly
help the tuning of the complicated Monte Carlo calculations used
to evaluate the development of the showers induced by cosmic rays
in the upper atmosphere. If extrapolations from our
parametrization model are correct this would imply that
$\sigma_{p-air}^{inel}$ should be smaller than usually considered,
which would have important consequences for development of high
energy cascades. Though it is quite clear that the statements of
this work are based on the assumption $\sigma_{tot}^{pp}$ =
 $\sigma_{tot}^{\bar{p}p}$ at energies in the range 546
$\leq  \sqrt s  \leq$ 1800 GeV, we claim that this {\sl
``enigma''} will be solved by the forthcoming new very high energy
colliders. The new  Relativistic Heavy Ion Collider (RHIC) which
has become operational at Brookhaven, should produce  a   value
of  $\sigma_{tot}^{pp}$ at  $\sqrt{s} =  500$ GeV very soon,
allowing  a comparison with the  $\sigma_{tot}^{\bar{p}p}$
measured value at $\sqrt{s} = 546$ GeV. Later on,  the LHC should
give us  a $\sigma_{tot}^{pp}$ value  at $\sqrt{s} =  14$ TeV. If
it  could be  run  at 2 TeV, a  direct  comparison between its
$\sigma_{tot}^{pp}$  value and the  $\sigma_{tot}^{\bar{p}p}$
found  at the Tevatron  could be done. Additionally, experiments
such as the HiRes \cite{hires} and the Auger Observatory
\cite{Auger} will bring new highlights to solve the puzzle  about
extrapolations to cosmic ray energies, i.e. whether the real
values of $\sigma_{tot}$ are described by parameterizations
consistent with a fast energy rise at high energies, which  we
may  called {\sl ``cosmic result''} or by a slower rise in energy,
the {\sl ``accelerators result''} as is the case presented in this
paper. We argue that instead of looking for a new physics based on
``exotic'' events to explain very fast rise in energy of
$\sigma_{tot}$, what we need is highly trustable data at
intermediate energies ($   2 \leq \sqrt{s}  \leq  15$ TeV) to use
them as  confident lever arms for extrapolations, to approach in
this way to a more accurate description of  $\sigma_{tot}$ at
cosmic ray energies.  In summary, extrapolations from accelerator
data cannot be disregarded to constraint cosmic ray estimations.

\appendix
\section{CONFIDENCE INTERVALS OF THE MODEL PREDICTION}
The ability of a statistical method  to reproduce data  of any physical quantity with high precision gives the pattern  for the
prediction of out of the range data.  In the context of
$pp$ and $\bar{p}p$ hadronic total  cross sections at very high
energies a great deal of work has been done out of the energy range of accelerators: predictions are usually compared to cosmic ray data producing a disagreement which explanation has also
been widely discussed in the literature.
Such comparison depends  critically  of a highly confident band
of  uncertainty for any parametrization model.  The validity
of any statistical method to predict a given physical quantity out of
the range values, ``prognosis'' (extrapolation) depends on its
precision to reproduce the employed data (interpolation).
A fundamental task  of any prediction method is
to minimize the error band of the predicted set of values.  In the specific
case of $\sigma_{tot}$ what is searched is to
obtain a prediction beyond the energy range of the employed data with the minimum of dispersion. In this context, popular
techniques are derived from the statistical theory known as
{\it Regression Analysis} either on the {\it multiple regression}
approach or on the simplest version, a {\it simple regression analysis},
both based on the method of {\it Least Squares}.
Among the important  indicators of any statistical method within
the frame of the present work there is the {\it confidence level}
 (C.L.) and the {\it confidence intervals} (e.g., \cite{Mont}).

For a given phenomenon characterized  by an independent variable (
x ) that generates a response variable ( y)  {\it regression
analysis} supplies a procedure to estimate the corresponding
statistics: to fit a set of data of the phenomenon in between the
known points (interpolation),  to {\em estimate}  $(\bar{y})$  the
mean value of (y), to predict a future value of (y)  beyond the
known points ({\em prognosis} out of the data range)  for a given
value of (x) ),  and to build a confidence interval around each of
the {\em estimated} or {\em prognosticated} values (for instance
\cite{Cramer,Eadie}). In its general form the dependent variable
(y) can be written as a function of ${\it k}$ independent
variables $x_1, x_2,....x_{\it k}$, where the variables $x_j$  may
represent powers of these variables, cross products of the
variables, or even to present a parametric dependence  of other
variable (see for instance \cite{M&S}).   Statistic techniques of
{\it regression analysis} are
 based on the minimization of the quadratic sum of data deviations with respect to the
employed mathematical model of prediction:  for a given distribution of {\it residuals} $R_i =y_{i} - \bar{y}_{i}$
 a full set of confidence indicators are generated, where  typical error bands (belts) for extrapolations
up to ${\it  m}$ steps beyond the last $n-esim$ experimental point  is given as
\begin{equation}
ErrorB = \bar{y}_{n+m} \pm t_{\delta/2}^{k-p} S_{d}
\end{equation}
where $S_{d}^{2}$ is the {\it variance}, defined as the square of the {\em standard deviation} (usually called the
standard error of estimate), $\bar{y}_{ n+m}$ is the corresponding {\it Central prediction} and
$t^{ \{ k-p \} }_{ \{ \delta /2 \} }$ denotes the Probability Density Function (p.d.f.) known as the Student's  $\{ \bf{t} \}$
-distribution for the $k$ values of the
independent variables with $p$ degrees of freedom, and  a confidence coefficient $\delta /2$. However, the previous generalization
does not incorporates effects related to the position of
the experimental values of (y) around the
employed Central Value, that is  the {\it correlation}
among {\it residuals} that generates the regression model \cite{M&S}.  In other words, it is assumed no interaction terms,
implying that each of the independent variables affects the
response $(y)$ independently of the other independent variables and
the error $\epsilon$ associated with any one $(y)$ value
is independent of the error associated with any other $(y)$ value.
Such an omission of the {\it residuals} correlation leads to
a notorious modification of the statistic estimators \cite{Cude}.

\subsection{The {\em {\em Forecasting}} Method}

The previous limitation, mentioned above, can be surmounted by
identifying the obtained data set $(x_{1i},x_{2i},x_{3i}......
x_{ki}; y_{i})$ with a time series, and  then to proceed an
evaluation of the correlation among consecutive {\it residuals}
(the so called {\it autocorrelation} procedure, from which the
simplest one is the {\it autoregressive 1st-order} model). The
{\em {\em {\em forecasting}}} statistic method is based on
autocorrelation models which allows for the evaluation of the
correlation among consecutive {\it residuals}, out and in the set
of data by means of an iterative process \cite{M&S}. In general,
this procedure modifies the fitting constants of the model, the
estimated variance of {\it residuals} and minimizes the width of
the error intervals for  interpolation or  prognosis
(extrapolation), increasing in this way the confidence level
$(CL)$ of predictions \cite{M&S}. To quantify the effect that has
the autocorrelation of {\it residuals} on the {\it regression}
model and associated estimators,  let us represent such a model by
a response variable  $y = E(y) + \epsilon$  where
\begin{equation}
E(y) = \beta_{0} + \beta_{1}x _{1} + \beta_{2}x_{2} + ........+\beta_{k}x_{k}
\end{equation}
represents the deterministic component of the proposed {\it
regression} model, $x_{1},x_{2},...,x_{k}$ are the independent
variables, that are known without error, and  may represent higher
order terms and even functions of variables as long as the
function do not contain unknown parameters,
$\beta_{1},\beta_{2},...., \beta_{k}$ are the unknown coefficients
to be determined by the least squares method, representing the
contribution of the independent variables $x_{i}$ and $\epsilon$
represents the random error component. For the evaluation of {\it
residuals} $R_{i} = y_{i}- E(y_{i})$ it is assumed they have a
{\it normal}  distribution which mean is zero and the variance is
constant.  In the autoregressive model of 1st-order each {\it
residual} $R_{i}$ is related with the previous one as
\begin{equation}
R_{i} = \phi R_{i-1} + {\it r_{i}}
\end{equation}
where $\phi$ is the autocorrelation constant among the {\it residuals} $(\mid \phi \mid < 1)$,  \cite{Full, Chat}, and ${\it r_{i}}$ in this case is a
Residual called white noise, uncorrelated of any other Residual component. The incorporation of the effect of autocorrelation
of {\it residuals} to the solution of the {\it regression} problem through equation (A2) leads to a modification of the {\it regression}
constants and the corresponding variance: using the set of data, the estimation of  the $k + 1$ regression constants of eq. (A2) and
the  constant $\phi$ are obtained from the  autocorrelation model according to the following interpolation equation for the
response variable $\hat{y}_{i}$:
\begin{equation}
\hat{y}_{i} = \hat{\beta}_{0} + \hat{\beta}_{1}x _{1,i} + \hat{\beta}_{2}x_{2,i} + ........\hat{\beta}_{k}x_{k,i}+ \hat{\phi}\hat{R}_{i-1}
\end{equation}
where $i= 1...n$, $x_{k,i}$ represents the independent variable
$x_{k}$ corresponding to the point $(i)$, the small hat indicates
the {\it estimated} values, that is they have been evaluated, and
the mean of the residual white noise has been taken as $\bar{\it
r}_{n+1}= 0$. The prediction of values beyond the n-esim data
value, beginning for instance with $y_{n+1}$, is then given as
\begin{equation}
y_{n+1} =\hat{\beta}_{0} + \hat{\beta}_{1}x _{1,n+1} + \hat{\beta}_{2}x_{2,n+1} + ........+\hat{\beta}_{k}x_{k,n+1}+ R_{n+1}
\end{equation}
where $x_{k,n+1}$ represents the independent variable $x_{k}$
corresponding to the point $(n+1)$. The forecast (prognostic) for
the response variable $\hat{y}_{n+1}$ is :
\begin{equation}
\hat{y}_{n+1} = \hat{\beta}_{0} + \hat{\beta}_{1}x _{1,n+1} + \hat{\beta}_{2}x_{2,n+1} + ........+\hat{\beta}_{k}x_{k,n+1}+ \hat{\phi}\hat{R}_{n}
\end{equation}
Similarly, for a posterior value $\hat{y}_{n+2}$,
\begin{equation}
\hat{y}_{n+2} = \hat{\beta}_{0} + \hat{\beta}_{1}x _{1,n+2} + \hat{\beta}_{2}x_{2,n+2} + ........+\hat{\beta}_{k}x_{k,n+2}+ {(\hat{\phi})}^{2}\hat{R}_{n}
\end{equation}
and so on successively, that is the {\em {\em forecasting}} of
values out of the range of data is an iterative process where
every new prognostic make use of the previous Residual.
Unfortunately, as we move more and more far from the last data
value,  the potential error increases due to possible changes in
the structure of the regression model, or changes in the variance
value as the prognostic are taking place. The later can be
surmounted by estimating the variance associated with each
prognostic ({\em {\em forecasting}} variance) through the
correlation constant $\phi$; therefore, according to the
autoregressive model of 1st-order within the data range  up to the
data $n$ we have a constant variance $S_{f}^{2}$, and for one step
out of the data range (i.e., $n+1$) the corresponding variance is
$S_{f,n+1}^{2} = S_{f}^{2}[1 + \phi^{2}]$, whereas for two steps
beyond $(n+2)$ the {\em {\em forecasting}} variance is
$S_{f,n+2}^{2} = S_{f}^{2}[1 + \phi^{2}+\phi^{4}]$ and for $m$
steps beyond the range $(n+m)$ the {\em {\em forecasting}}
variance is $S_{f,n+m}^{2} = S_{f}^{2}[1 + \phi^{2} + \phi^{4}
+....+ \phi^{2(m-1)}]$. On this basis, for a prediction interval
with a confidence of  $100(1-\delta)\%$ and a type $t-student$
distribution, the amplitude of the prognostic intervals for
$(n+m)$ steps beyond the $n-esim$ data value is given in
\cite{M&S}, as:
\begin{equation}
ErrorB = \hat{y}_{n+m} \pm t_{\delta/2}^{k-p} \sqrt{S_{f}^{2}[1 + \phi^{2} + \phi^{4} +....+ \phi^{2(m-1)}]}
\end{equation}
Due to the incorporation of the autocorrelation of {\it
residuals}, the error bands evaluated in this way gives a higher
Confidence Level than other methods of Regressive Analysis which
ignore this effect \cite{M&S},  what is translated in a width
decrease of the prediction intervals as a consequence of the
decrease of the estimated variance(e.g., Table 9.7 in \cite{M&S}).
It must be emphasized  that technically the meaning of the error
bars in different methods derived from {\it  regression analysis}
is exactly the same , since they quantify the level of confidence
(C.L.), that is, they express the probability that the {\it ``true
answer''} can be placed with $100(1-\delta)\%$ of probability
within the interval of aleatory measurements, or predictions, when
other measurement is done under the same conditions. However, even
if the concept is the same , every method predicts a different
value of (C.L.): in the particular case of  the {\em {\em {\em
forecasting}}}, the essential point  is that it uses the
additional factor of the autocorrelation among {\it residuals},
what improves (1) the {\it Central prognostic} value and (2) the
amplitude of the confidence intervals (error bars), whose
narrowing  is translated precisely in a high level of confidence
\cite{M&S}.

\subsection{Matrix Approach in  Regression Analysis}
This {\em {\em {\em forecasting}}} method is based on the {\it
multiple  regression} theory and consists in  determining a
prediction equation for a quantity $y$ (dependent variable), that
in turns depends on $k$ independent variables ($x_{i}$), that is
\begin{equation}
E(y) =  \sum_{i=0}^{k} \beta_{i} f_{i} (x_{i})
\end{equation}
(with  $ f_{o}(x_{o}) = 1$) ,  where $f_{i}$ are arbitrary
functions of $x_{i}$, and $\beta_{i}$ are the fitting constants.
In the generalized version the variable $x_{i}$ may depend on
other parameters, i.e., $x_{i} = x_{i}(s,t,..)$.  Therefore, the
application of a multiple regression model to a given problem
leads to  trace a system of ·$n$ equations with $n$ incognitos, so
that its solution is better obtained through a matrix formalism.
Denoting with $Y$ the matrix of $(n \times 1)$-dimension of the
dependent variables  and with $X$ the matrix of $[ n \times (k +
1)]$-dimension of the $k$ independent variables,  the row  ${1,
x_{11}, x_{12},.....,x_{1k}}$ multiplied by the column matrix of
the $\beta 's$ determines the value $y_{1}$ of the dependent
variable, the row  ${1, x_{21},  x_{22}, .....,x_{2k}}$ multiplied
by the column matrix of the $\beta 's$ determines the value
$y_{2}$ and so on.

$$X = \left(  \begin{array}{ccccc}
1&  x_{11}&  x_{12}&  \cdots&  x_{1k}\\
1&  x_{21}&  x_{22}&  \cdots&  x_{2k}\\
\vdots&  \vdots&  \vdots&  \ddots&  \vdots\\
1&  x_{n1}&  x_{n2}&  \cdots&  x_{nk}
\end{array}  \right); Y = \left(  \begin{array}{c}
y_{1}\\
y_{2}\\
\vdots\\
y_{n}
\end{array}  \right); \mathcal{B} = \left(  \begin{array}{c}
\beta_{0}\\
\beta_{1}\\
\vdots\\
\beta_{k}
 \end{array}  \right)$$
The  variables contained in the matrixes $X$, $Y$ can be related
by the matrix equation $Y  =  X\mathcal{B}$ , which is the matrix
expression of   the prediction  equation (A9). The $[ (k+1) \times
1 ]$-dimension matrix $\mathcal{B}$ contains the values of the
constants $\beta_{i}$ needed to write in explicit form the
prediction equation (A9). The $\beta's$ constants can be
determined by the least squares method  \cite{M&S} through the
condition
\begin{equation}
\sum_{j=1}^{n}  [ y_{j} - E(y_{j})]^{2} = \sum_{j=1}^{n} R_{j}^{2} = {\it  minimum}
\end{equation}
where $y_{j}$ is the $j-esim$ measurement of the response variable and $E(y_{j})$ is the estimated Central Value with Eq. (A9).
The condition (A10) is satisfied when $\frac{\partial}{\partial \beta_{i}}\sum_{j=1}^{n} R_{j}^{2}$ = $0$,  $(i = 1......k)$ which leads to
a system of $n$ equations with $k(=n)$ unknowns. This system in matrix formalism can be written (e.g., \cite{Basil}) as
\begin{equation}
(X^{t} X)\hat{\mathcal{B}} =  X^{t}Y
\end{equation}
where $ X^{t}$ denotes the transposed matrix of $X$  and $\hat{\mathcal{B}}$ is the matrix of the expected values of the
$\beta's$. From (A11) we obtain the solution  of eq. (A10):
\begin{equation}
\hat{\mathcal{B}} = (X^{t} X)^{-1}X^{t}Y
\end{equation}
where $(X^{t} X)^{-1}$ denotes the inverse matrix of $ X^{t} X $.
Essentially this equation minimizes the quadratic sum of the deviations of points $y_{j}$ with respect
to the prediction equation  (A9)  (\cite{M&S} p. 783).
With the previous  matrixes several statistical estimators
are easily determined, such as the Sum of Square Errors (SSE)
\begin{equation}
SSE = Y^{t}Y- \hat{\mathcal{B}}^{t}(X^{t}Y)
\end{equation}
 and the variance required to evaluate the confidence intervals is:
\begin{equation}
 \it{S}_{f}^{2} = \frac {SSE}{[n - (k+1) ]}
\end{equation}
where the denominator defines the number of degrees of freedom  for errors,
given by the number  of $\beta_{i}$ -
parameters.
Once the {\it Central} values are known,  we then evaluate
the confidence interval
for a particular value of the response variable,  $y_{p}$,
using  the matrix of the particular values of the independent variables
which determine the estimated value of $y_{p}$. Such a matrix,
namely  $\mathcal{A}$,  denotes  the column-matrix of $(k+1) \times 1$
dimension,  which elements $\{ 1,  x_{1p}, x_{2p}, \ldots, x_{kp}\}$
correspond to the numerical values of the $\beta _{i}$
appearing in equation $(A9)$. Therefore, according to the associated
statistic the confidence interval for
prediction within the range of data is determined as (\cite{M&S} p. 795):
 \begin{equation}
InterpB = \hat{y} \pm t^{n-p}_{\delta/2} \sqrt{ \it{S}_{f}^{2}A^{t}(X^{t}X)^{-1} A }
\end{equation}
and for  prognosis out  of  the  data  range as (\cite{M&S} p. 800):
\begin{equation}
ExtrapB = \hat{y} \pm t^{n-p}_{\delta/2} \sqrt{ \it{S}_{f}^2 \left[ 1+A^{t} \left( X^{t}X \right)^{-1}A \right] }
\end{equation}
Here $\hat{y}$ denotes the central prediction, $A^{t}$ is the transposed matrix of $A$.
$InterpB(+)$, $ExtrapB(+)$ and $InterpB(-)$, $ExtrapB(-)$ denote the corresponding
Upper and Lower bounds respectively.   $t^{ \{ n-p \} }_{ \{ \delta /2 \} }$ denotes Student's  $\{ \bf{t} \}$-distribution for the
 $n$ values of the independent variables with $p$ degrees of freedom. Estimation have been done with  a precision of
$100(1-\delta)\%$, assuming  ${\delta/2}=0.025$,  which corresponds to  a value of $95\%.$

\end{document}